\documentclass[
 reprint,
 amsmath,amssymb,
 aps,
pra,
floatfix
rsi]{revtex4-2}

\usepackage{graphicx}
\usepackage{dcolumn}
\usepackage{bm}
\usepackage{amsmath}
\usepackage{physics}

\newcommand{\up}{\uparrow}
\newcommand{\dow}{\downarrow}
\newcommand{\bfp}{{\bf p}}

\begin{document}

\preprint{APS/123-QED}

\title{Three-body interactions in Rabi-coupled Bose gases: a perturbative approach}
\author{S. Tiengo}
\author{R. Eid}
\author{T. Bourdel}
\thanks{thomas.bourdel@institutoptique.fr}
\affiliation{%
 Université Paris-Saclay, Institut d'Optique Graduate School,\\
 CNRS, Laboratoire Charles Fabry, 91127 Palaiseau, France 
}%

\date{\today}

\begin{abstract}
In ultracold atomic gases, radio-frequency coupling between two spin states can lead to atoms being in a stable coherent superposition of the two states (dressed states). When the two-body interactions (scattering lengths) are different between the two states, it offers the possibility not only to control the interatomic interaction but also to modify the equation of states with emerging three-body terms with different signs and scalings with the parameters. We derive these terms in a unified perturbation framework and show that they correspond to genuine three-body processes found at different orders in perturbation theory. Our work justifies the introduction of contact three-body interactions in dressed-state systems when the spin degree of freedom is integrated out.  
\end{abstract}

\maketitle

\section{Introduction}
Ultracold quantum gases are model many-body systems that permit one to access a variety of physical phenomena, such as superfluidity or phase transitions \cite{Bloch2008}. A key simplification comes from the fact that they are dilute systems with a dominating two-body interaction that, in addition, can be modeled by a contact interaction \cite{Varenna98}. In this context, the scattering length $a$ is the only quantity that fully characterizes the interaction strength. Importantly, the scattering lengths can be modified through diffusion (Fano-Feshbach) resonances by tuning the magnetic field \cite{Chin2010}. This degree of control and the versatility of the external potential allows for versatile and quantitative studies (or quantum simulations) of complex many-body problems such as the BEC-BCS transition for fermions \cite{Regal2004, Bartenstein2004, Zwierlein2004, Bourdel2004}, the Mott-transition \cite{Greiner2002}, the Berenzinski-Kosterlitz-Thouless superfluid transition in 2D \cite{Hadzibabic2006} or the generalized hydrodynamics in 1D \cite{Schemmer2019} for bosons. 

Physics can be made even richer when mixtures are used such that different two-body interactions may coexist \cite{Varenna25}. For bosons, this has led to the observation of phase separation for repulsive interspecies interaction \cite{Hall1998} and quantum droplets for attractive interspecies interaction \cite{Petrov2015, Cabrera2018}. Coupling between two spin states (for example, using a radio-frequency field) adds another degree of freedom. Competition between spin interaction and coupling can lead to bistability \cite{Zibold2010} or ferromagnetism phenomena \cite{Cominotti2024} in the spin degree of freedom. Coherent coupling also leads to an alternative way of controlling the two-body interaction \cite{Sanz2024}. Indeed, the effective interaction then varies with the spin composition of the dressed state, which can be controlled by the detuning of the coupling field. In this configuration, it was also shown that three-body interaction terms either positive or negative emerge in the equation of state and may play a dominant role in the dynamics of Bose-Einstein condensates. These terms were found to appear either from mean-field shift of the energy difference between the two spin states \cite{Hammond2022}, or in complex calculations beyond the mean field in the limit of large coupling \cite{Lavoine2021}. These experimental demonstrations of controllable three-body interactions without strong enhancement of losses in ultra-cold dilute gases came in a context where such interactions have been considered in numerous theoretical works \cite{Wu59, Gammal00, Gammal00b, Kohler02, Bulgac02, Tan08, Kumar10,Crosta12, Petrov2014, Petrov14lattice, Zloshchastiev17,Killip17, Incao18, Sekino2028, Mestrom19, Zwerger19}.

In this paper, we perform perturbation theory as a function of the interaction strengths and up to the third order. We recover the three-body interaction terms for a Bose gas with coherent coupling between two spin states in a unified framework at different orders in perturbation. We first treat the problem of a homogeneous condensate in the dressed state and calculate the modification of its ground-state energy. This calculation directly links to previous results on the equation of states \cite{Lavoine2021, Hammond2022}. Second, few-body scattering processes \cite{Petrov2014} are considered and calculated. Our results justify the introduction of a contact three-body interaction term in an effective Hamiltonian in the reduced Hilbert space, in which the spin excitations are no longer considered. Finally, a generalization of the calculations to one dimension (1D) is presented.

\section{Hamiltonian in the dressed-state basis}
In the second quantization formalism, the single particle and the interaction Hamiltonians $H_0$ and $H_\textrm{int}$ in a coupled two-component homogeneous Bose system of volume $V$ read: 
\begin{widetext}
\begin{gather}
H_0 =\sum_{\bf p}\left(\frac{\bfp^2}{2m}+ \frac{\hbar\delta}{2}\right) a^\dagger _{\up \bfp} a _{\up \bfp}+\sum_{\bf p}\left(\frac{\bfp^2}{2m} - \frac{\hbar\delta}{2}\right) a^\dagger _{\downarrow \bfp} a _{\downarrow \bfp}+\sum_{\bf p} \frac{\hbar \Omega}{2} (a^\dagger _{\up \bfp} a _{\downarrow \bfp}+a^\dagger _{\downarrow \bfp} a _{\up \bfp}),\\
H_\textrm{int}=\frac{1}{2V}\sum_{\bfp_1+\bfp_2=\bfp_3+\bfp_4}(g_{\up \up}a^\dagger _{\up \bfp_3}a^\dagger _{\up \bfp_4}a _{\up \bfp_1}a _{\up \bfp_2}+g_{\downarrow \downarrow}a^\dagger _{\downarrow \bfp_3}a^\dagger _{\downarrow \bfp_4}a _{\downarrow \bfp_1}a _{\downarrow \bfp_2}
    +2g_{\up \downarrow}a^\dagger _{\up \bfp_3}a^\dagger _{\downarrow \bfp_4}a _{\up \bfp_1}a_{\downarrow \bfp_2}),
\end{gather}
\end{widetext}
where $m$ is the atomic mass, $\hbar$ the reduced Planck constant, $\delta$ the coupling detuning, and $\Omega$ the Rabi frequency of the coupling between the two spin states. $a^\dagger _{\sigma \bfp}$ and $a _{\sigma \bfp}$ are the creation and annihilation operators of a particle with a certain spin $\sigma$ in the $\{ \uparrow, \downarrow\}$ basis and momentum $\bfp$. In the interaction Hamiltonian appear the three coupling constants $g_{\sigma\sigma'} = 4\pi\hbar^2 a_{\sigma\sigma'}/m$ for the $\sigma-\sigma'$ contact interactions with scattering lengths $a_{\sigma\sigma'}$. 

In order to perform perturbation theory, we first diagonalize the single particle Hamiltonian $H_0$ by a rotation to the dressed-state basis $\ket{+}$, $\ket{-}$, thus introducing the related operators
\begin{gather}
a_{+\bfp}=\cos \theta a_{\downarrow \bfp}+\sin \theta a_{\up \bfp} \notag,\\
a_{-\bfp}=\sin \theta a_{\downarrow \bfp}-\cos \theta a_{\up \bfp},
\end{gather}
where $\theta$ is the mixing angle such that $\textrm{cotan} (2 \theta)= \delta/\Omega$. In this description, $\theta$ evolves from 0 to $\pi/2$ and is $\pi/4$  at resonance (that is, for $\delta=0$).
In this rotated basis $H_0$ reads:
\begin{align}
H_0 =\sum_p\frac{\bfp^2}{2m} (a^\dagger _{+ \bfp} a _{+ \bfp}+a^\dagger _{- \bfp} a _{- \bfp})\nonumber\\+\sum_p \frac{\hbar \tilde{\Omega}}{2} (a^\dagger _{+ \bfp} a _{+ \bfp}-a^\dagger _{- \bfp} a _{- \bfp}),
\end{align}
where $\tilde{\Omega}=\sqrt{\Omega^2+\delta^2}$. In the same basis, the interaction Hamiltonian \cite{Search01} is
\begin{align}
\label{eqn:Hint}
&H_\textrm{int}=\frac{1}{2V}\sum_{\bfp_1+\bfp_2=\bfp_3+\bfp_4}\notag\\\Big(& g_1 a^\dagger _{+ \bfp_3}a^\dagger_{+ \bfp_4}a _{+ \bfp_1}a_{+ \bfp_2}\notag\\
&+ g_2 a^\dagger _{- \bfp_3}a^\dagger_{- \bfp_4}a _{- \bfp_1}a_{- \bfp_2} \notag\\
&+ g_3 a^\dagger _{+ \bfp_3}a^\dagger_{- \bfp_4}a _{+ \bfp_1}a_{- \bfp_2} \notag\\
&+ g_4 (a^\dagger _{+ \bfp_3}a^\dagger_{+ \bfp_4}a _{- \bfp_1}a_{- \bfp_2}+a^\dagger _{- \bfp_3}a^\dagger_{- \bfp_4}a _{+ \bfp_1}a_{+ \bfp_2}) \notag\\
&+ g_5 (a^\dagger _{+ \bfp_3}a^\dagger_{+ \bfp_4}a _{+ \bfp_1}a_{- \bfp_2}+a^\dagger _{+ \bfp_3}a^\dagger_{- \bfp_4}a _{+ \bfp_1}a_{+ \bfp_2}) \notag\\
&+ g_6 (a^\dagger _{+ \bfp_3}a^\dagger_{- \bfp_4}a _{- \bfp_1}a_{- \bfp_2}+a^\dagger _{- \bfp_3}a^\dagger_{- \bfp_4}a _{+ \bfp_1}a_{- \bfp_2})\Big).
\end{align}
The first three terms of the Hamiltonian correspond to elastic collisions between two atoms in the dressed states, with, respectively, $g_1$, $g_2$, $g_3$ as the interaction strength coefficients. The last three terms with coupling constants $g_3$, $g_4$ and $g_5$, instead describe scattering events that do not preserve the particle dressed state (inelastic collisions). 

The dependence of the coupling coefficients on the mixing angle (or equivalently on the detuning $\delta/\Omega$) is:
\begin{align}
& g_1=g_{\downarrow \downarrow} \cos^4 \theta+g_{\up \up} \sin^4 \theta +\frac{1}{2} g_{\up \downarrow}\sin^2 2\theta\notag\\
& g_2=g_{\downarrow \downarrow} \sin^4 \theta+g_{\up \up} \cos^4 \theta+\frac{1}{2} g_{\up \downarrow} \sin^2 2\theta \notag\\
& g_3=(g_{\downarrow \downarrow}+g_{\up \up}) \sin^2 2\theta +2 g_{\up \downarrow} \cos^2 2\theta \notag\\
&g_4=\frac{1}{4}(g_{\downarrow \downarrow}+g_{\up \up}-2g_{\up \downarrow})\sin^2 2\theta \notag\\
& g_5=\sin 2\theta (g_{\up \up} \sin^2 \theta-g_{\downarrow \downarrow} \cos^2 \theta+g_{\up \downarrow}\cos 2\theta) \notag\\
& g_6=\sin 2\theta (g_{\up \up} \cos^2 \theta-g_{\downarrow\downarrow} \sin^2 \theta-g_{\up \downarrow}\cos 2\theta).
\end{align}
\begin{figure}[htbp!]
\includegraphics[width=\columnwidth]{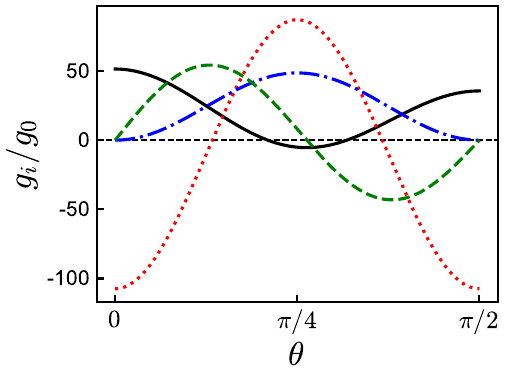}
\caption{\label{fig:gi} $g_2$ (black solid line), $g_3$ (red dotted line), $g_4$ (blue dash-dotted line), $g_6$ (green dashed line) as a function of the mixing angle $\theta$ and normalizes to $g_0=4\pi\hbar^2a_0/m$, where $a_0$ is the atomic Bohr radius. $a_{\up\up}=51.5\,a_0$, $a_{\dow\dow}=35.6\,a_0$, $a_{\up \dow}=-53.7\,a_0$. 
}
\end{figure}
 The precise description of higher-order interaction processes is especially relevant in conditions where the usual two-body collisions is reduced thanks to the compensation between positive and repulsive bare coupling constants. In such a situation, higher-order interactions will play an important role. It occurs in particular when the intraspecies coupling constant $g_{\up \up}$ and $g_{\dow \dow}$ are positive whereas $g_{\up \dow}$ is negative, conditions that can be met in potassium 39 \cite{D'Errico2007}. As an example, the coupling constants $g_2$, $g_3$, $g_4$, $g_6$ are plotted for the second and third lowest states of $^{39}K$ at a magnetic field of 55.6\,G in Fig.\ref{fig:gi}. If $\theta=0$ or $\theta=\pi/2$ (i.e. $\delta/\Omega \rightarrow \pm \infty$), we find only elastic collisions and the original Hamiltonian in the non-rotated internal state basis. In the vicinity of the resonance $\delta \approx \pi/4$, $\ket{-}$-$\ket{-}$ interactions are reduced as $g_2$ is close to zero. Note that $g_6 \propto \frac{\partial g_2}{\partial \theta}$ such that $g_6$ vanishes when $g_2$ is minimal.

\section{Perturbative expansion of the condensate energy}
We first perform perturbation theory in order to calculate the ground state energy corresponding to a Bose-Einstein condensate in state $\ket{-, {\bf p=0}}$. The unperturbed ground state is $\ket{\psi_0}=\frac{1}{\sqrt{N!}}(a^\dagger_{-,\bfp=0})^N\ket{0}$, where $N$ is the atom number. At the first order in perturbation theory, the correction to the energy is given by the expectation value of $H_\textrm{int}$ in the unperturbed ground state
\begin{align}
E^{(1)}=\bra{\psi_0}H_\textrm{int} \ket{{\psi_0}}. 
\end{align}
Among all the terms of $H_\textrm{int}$, only the term multiplied by the coupling constant $g_2$ allows two colliding atoms to remain in the dressed state $\ket{-}$. There is therefore a single term in the sum that contributes at the first order
\begin{align}
\label{eqn:1-order}
E^{(1)}&=\frac{g_2}{2V} \bra{\psi_0} a^\dagger _{-,\bfp=0}a^\dagger_{-,\bfp=0}a _{-,\bfp=0}a_{-,\bfp=0} \ket{{\psi_0}}\nonumber\\
&=g_2\frac{N(N-1)}{2V} \simeq g_2\frac{N^2}{2V}. 
\end{align}
The first order correction (eq.\,\ref{eqn:1-order}) corresponds to the standard mean-field interaction energy with a coupling constant $g_2$ that varies with the detuning $\delta/\Omega$. 

The energy correction at the second order is given by
\begin{align}
E^{(2)}=\sum_i\frac{\bra{\psi_0}H_\textrm{int} \ket{{\psi_i}}\bra{\psi_i}H_\textrm{int} \ket{{\psi_0}}}{E_0-E_i}, 
\end{align}
where the sum is taken over all possible intermediate states $\ket{\psi_i}\ne\ket{\psi_0}$.
We can recognize that the terms of $H_\textrm{int}$ that contribute to the second order are those multiplied by the coupling constants $g_2$, $g_4$, and $g_6$. 
At the second order the energy correction is given by
\begin{align}
E^{(2)}=-\frac{g_2^2N^2}{2V^2}\sum_{\bfp\ne0} \frac{m}{\bfp^2}-\frac{g_4^2N^2}{2V^2}\sum_{\bfp\ne0} \frac{1}{2\hbar \tilde{\Omega} +\bfp^2/m}\nonumber\\-\frac{g_6^2 N^2}{4V^2}\sum_{\bfp\ne0} \frac{1}{\hbar \tilde{\Omega}+\bfp^2/m}-\frac{g_6^2N^3}{4V^2\hbar\tilde{\Omega}}
\end{align}
Note that we have omitted the terms that are not extensive ($\propto N$) in the thermodynamic limit (i.e. for large $N$ and $V$, with constant density $n=N/V$). The last term originates from the term with all momenta equal to 0 and scales as $N^3$ because there are three $a_{-0}$ or $a^\dagger_{-0}$ in the corresponding Hamiltonian term. It corresponds to the negative three-body energy contribution calculated at the mean field level, taking into account a density-dependent shift of the detuning \cite{Hammond2022}. 

The first three terms are divergent at large ${\bf p}$. This is a consequence of the unphysical nature of the contact potential in 3D, which needs to be regularized, for example, by adding a momentum cutoff in $H_\textrm{int}$. An alternative way is to renormalize the coupling constant, which is equivalent to removing the divergent parts in the sums \cite{Pethick2008}. After such a procedure, the first term disappears, whereas the two other terms are renormalized 
\begin{align}
E^{(2)}=\frac{(2\sqrt{2}g_4^2 +g_6^2)N^2\sqrt{\hbar\tilde{\Omega}}}{2V}
\frac{m^{3/2}}{8\pi\hbar^3}-\frac{g_6^2N^3}{4V^2\hbar\tilde{\Omega}}.
\end{align}
In the specific case, where $g_2$ is minimal and zero ($g_6=0$), we exactly recover the renormalization of the two-body interaction due to beyond-mean-field effects calculated from a Bogoliubov type approach in the limit of large coupling as compared to the interaction energy $\hbar\Omega\gg g N/V$ \cite{Lavoine2021}. 

Let us now consider third-order perturbation theory. Generally, it reads:
\begin{align}
E^{(3)}=\sum_i\sum_j\frac{{\bra{\psi_0}}H_\textrm{int} \ket{{\psi_i}}
\bra{\psi_i}H_\textrm{int} \ket{{\psi_j}}
\bra{\psi_j}H_\textrm{int} \ket{{\psi_0}}}
{(E_i-E_0)(E_j-E_0)}.
\end{align}
In principle, the sum includes terms with $g_4$ or $g_6$. Since, if $g_6$ is non zero, there is already a three-body contribution at the second-order of perturbation, we focus on the case where $g_6=0$. In that case, 
the only non-zero terms are of the form
\begin{align}
(g_4a^\dagger _{- 0}a^\dagger_{- 0}a _{+ \bfp_1}a_{+ (-\bfp_1)})(g_3 a^\dagger _{+ \bfp_2}a^\dagger_{- 0}a _{+ \bfp_2}a_{- 0})\nonumber\\(g_4a^\dagger _{+ \bfp_3}a^\dagger_{+ (-\bfp_3)}a _{- 0}a_{- 0}),
\end{align}
where the $\bfp=0$ are imposed by the fact that there is no atoms in $\ket{-, \bfp \ne 0}$ in $\ket{\psi_0}$. 

 The specific case of $\bfp_1=\bfp_2=\bfp_3=0$ can be neglected in the thermodynamic limit as it scales as $(N/V)^3$. For $\bfp_1\ne 0$, $\bfp_2$ and $\bfp_3$ have to take the same value up to a sign. There are thus 4 equivalent terms and
 \begin{align}
E^{(3)}&=\frac{g_4^2g_3N^3}{2V^3}\sum_{\bfp \ne 0} \frac{1}{(2\hbar\tilde{\Omega}+\bfp^2/m)^2}\nonumber\\
E^{(3)}&=\frac{g_4^2g_3n^3}{16\pi\sqrt{2\hbar\tilde{\Omega}}}\frac{Vm^{3/2}}{\hbar^3}.
 \end{align}
Again, it is exactly the positive value found in the Bogoliubov approach in the limit of large coupling $\hbar\Omega\gg g n$ \cite{Lavoine2021}. The pertubation theory permits us to find in a unified framework two-body and three-body interaction terms in the equation of state that were previously found in mean-field and beyond mean-field approaches. However, the perturbative expansion is less general than the Bogoliubov approach; it does not resolve the problem when $\hbar \Omega$ is not large compared to $gn$. In particular, perturbation theory cannot recover the nonanalytical Lee-Huang-Yang $N^{5/2}$ energy scaling for vanishing $\Omega$ \cite{Petrov2015}. 

\section{Three-body scattering in perturbation theory}
\begin{figure}
\includegraphics[width=0.7\columnwidth]{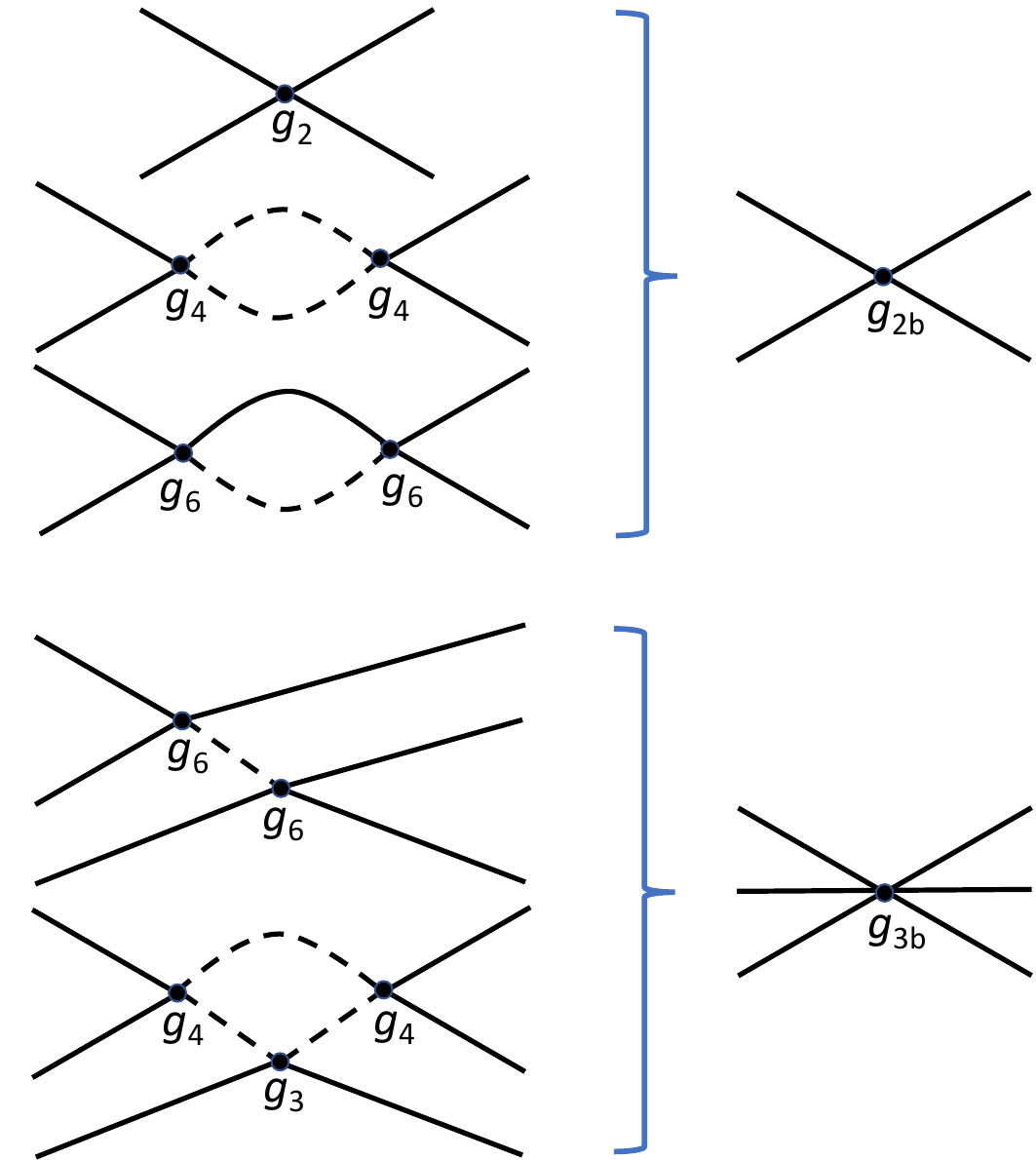}
\caption{\label{fig:Feynman} Feynman diagrams corresponding to the possible scattering events. The plain and dashed lines represent the $\ket{-}$ and $\ket{+}$ state, respectively. The three first diagrams correspond to two-body scattering at first and second order in perturbation theory and contributes to $g_{2b}$ in the effective interaction Hamiltonian. The two last diagrams correspond to three-body scattering at the second and third-order and contributes to $g_{3b}$ in the effective interaction Hamiltonian.
}
\end{figure}

The above calculation shows that the two- and three- body terms in the equation of state of a coherently dressed condensate can be found in perturbation theory. In this section, we show that they actually correspond to few-particle collisions, whose strength can be deduced from perturbative scattering theory. The corresponding Feynman diagrams are shown in Fig.\,\ref{fig:Feynman}. 

Let us first consider two-body elastic collisions from the initial state $\ket{\psi_i^{2b}}=a^\dagger_{-p_1}a^\dagger_{-p_2}\ket{0}$ to the final
state $\ket{\psi_f^{2b}}=a^\dagger_{-p_3}a^\dagger_{-p_4}\ket{0}$. Scattering theory tells that we should consider the matrix elements $\bra{\psi_f^{2b}}T\ket{\psi_i^{2b}}$, where $T$ is called the scattering matrix and can be expressed as
\begin{align}
T=H_\textrm{int}+H_\textrm{int}G_0H_\textrm{int}+H_\textrm{int}G_0H_\textrm{int}G_0H_\textrm{int}\nonumber\\+H_\textrm{int}G_0H_\textrm{int}G_0H_\textrm{int}G_0H_\textrm{int}+...., \\\textrm{where }
G_0=\frac{1}{H_0-E_0},
\end{align}
a form that is suited in order to apply perturbation theory. 

If we restrict to the first order (i.e. the first diagram in Fig.\,\ref{fig:Feynman} or Born approximation), only the $g_2$ term of the Hamiltonien contributes, and the matrix element has the same non-zero value for any value of the momenta that ensure total momentum conservation. It corresponds to the usual two-body $s$-wave scattering amplitude that leads to an isotropic scattering halo.  The second-order terms include as before contributions originating from $g_2$, $g_4$, and $g_6$, and the sums need to be regularized at high momenta with the usual renormalization of the coupling constant. Remarkably, for collisions at low momenta, i.e. such that $p^2/m \ll \hbar \Omega$, one can neglect the corresponding kinetic terms in the denominator such that the exact same type of terms as in the previous energy calculation appears. We thus find two additional terms up to the second-order in perturbation (corresponding to the second and third diagrams in Fig\,\ref{fig:Feynman})
\begin{align}
\bra{\psi_f^{2b}}T\ket{\psi_i^{2b}}\approx 2g_2/V+\frac{(2\sqrt{2}g_4^2+g_6^2)\sqrt{\hbar \tilde{\Omega}}}{4\pi V}.
\end{align}
This result can be simply interpreted as a modification of the scattering length due to the coupling. This correction can also be found from an exact calculation of two-body scattering in the presence of coupling (see supplemental material in \cite{Lavoine2021})

We now consider three particles in the $\ket{-}$ dressed state with momenta $p_1$, $p_2$, $p_3$ and ask the question if they can elastically scatter to other momenta $p_1'$, $p_2'$, $p_3'$. We thus consider $\bra{-p_1', -p_2', -p_3'}T\ket{-p_1, -p_2, -p_3}$. For simplicity, we consider the case where all momenta are different. At the second order in perturbation, the only terms in $H_\textrm{int}$ that allow us to change the three momenta are $g_2^2$ and $g_6^2$ because they need to annihilate two atoms in $\ket{-}$ and create at least one atom in $\ket{-}$. The $g_2^2$ term disappears with the renormalization of the coupling constant. The term $\propto g_6^2$ (fourth diagram in Fig.\,\ref{fig:Feynman}) first annihilates two $\ket{-}$ particles among the three, creates one in $\ket{+}$ and one in $\ket{-}$ with one of the final momenta, second annihilates the last initial $\ket{-}$ particle and the $\ket{+}$ particle, and creates the 2 other $\ket{-}$ particles with the final momenta. There are 9 possible intermediate states. If all momenta are small $p^2/m \ll \hbar \Omega$, the matrix element reduces to 
\begin{align}
\bra{\psi_f^{3b}}H_\textrm{int}G_0H_\textrm{int}\ket{\psi_i^{3b}}=\frac{9g_6^2}{V^2\hbar\tilde{\Omega}}
\end{align}
with no momentum dependence.

At the third order in perturbation theory, there are several possible terms, including $g_4$ or $g_6$. Similarly to before, we consider the case $g_6=0$, so that there is no three-body scattering at the second order. In that case, there is a single possible diagram (the fifth in Fig.\,\ref{fig:Feynman}). There remains one integration over a momentum which is not diverging and we find (after careful counting of the number of terms that contributes)
\begin{align}
\bra{\psi_f^{3b}}H_\textrm{int}G_0H_\textrm{int}G_0H_\textrm{int}\ket{\psi_i^{3b}}=\frac{9g_4^2g_3}{4\pi V^2\sqrt{2\hbar\tilde{\Omega}}}.
\end{align}
Interestingly, for symmetric conditions ($g_{\up\up}=g_{\dow\dow}$ and $\delta=0$), the latter term can was previously found in a exact three-body calculation \cite{Petrov2014}.

The above results show that two- and three-body interactions are also found in few-body scattering with the initial Hamiltonian. In the perturbative framework, the nature of the three-body interactions is clear. They correspond to higher-order scattering with intermediate states that have a different spin. 

Given the success of the perturbative expansion, it is tempting to continue the expansion to higher orders. However, we expect that the fourth order will not be accurate as corrections from short-range physics are then expected \cite{Wu59, Mestrom19}. Moreover, three-body recombination losses typically scale as $a^4$ \cite{Braaten2006} and also have to be considered at this level of perturbative expansion. 

\section{Effective Hamiltonian}
The above perturbative results justify an effective interaction Hamiltonian $H_{\textrm{eff}}$ in an Hilbert space reduced to the dressed state $\ket{-}$ with contact three-body interaction
\begin{align}
H_{\textrm{eff}}=&\frac{g_\textrm{2b}}{2V} \sum_{\bf p_1'+ p_2'= p_1+ p_2} 
a^\dagger_{p_2'}a^\dagger_{p_1'}a_{p_2}a_{p_1}\\+&\frac{g_\textrm{3b}}{3V^2} \sum_{\bf p_1'+ p_2'+ p_3'= p_1+ p_2+ p_3} 
a^\dagger_{p_3'}a^\dagger_{p_2'}a^\dagger_{p_1'}a_{p_3}a_{p_2}a_{p_1},
\end{align}
where $g_\textrm{2b}=V\bra{\psi_f^{2b}}T\ket{\psi_i^{2b}}/2$ and $g_\textrm{3b}=V^2\bra{\psi_f^{3b}}T\ket{\psi_i^{3b}}/12$. The numerical factors come from the number of permutations of the different momenta, and the corresponding two-body and three-body diagrams are shown on the right of Fig.\,\ref{fig:Feynman}. 
This effective Hamiltonian is constructed to give the same matrix elements as the ones found previously, thus leading to the same low-energy physics as the real system with the dressing field and possible virtual spin excitations. It also leads to previously calculated ground-state energy shifts, when applied to a condensate in the $\ket{-}$ state in the Born approximation. Higher-order terms originating from the effective Hamiltonian can be neglected at the level of accuracy of the initial third-order perturbation theory.

\section{Generalization to 1D systems}
In an elongated trap, the confinement may be such that the radial excitations of the gas correspond to the highest energy scale of the problem. In the simple case of a radial harmonic trap with frequency  $\omega_\perp/2\pi$, it requires $g_{\sigma\sigma'}n \ll \hbar \omega_\perp$ as well as $\Omega \ll \omega_\perp$. In the absence of copupling ($\Omega=0$), 1D quantum droplets have been predicted \cite{Petrov2016}. With coupling, 1D perturbation calculations can be done similarly to the 3D ones. The only difference is that the sums over the momenta are 1D rather than 3D. It is actually simpler because there is no divergence, and no renormalization of the coupling constants is necessary. The 1D coupling constants for a harmonic trap can be approximated by $g_{\sigma\sigma'}^\textrm{1D}=2\hbar\omega_\perp a_{{\sigma\sigma'}}$ in the limit $a_{{\sigma\sigma'}}\ll\sqrt{\hbar/m\omega_\perp}$ \cite{Olshanii1998}. 

The three-body interaction arising at the second-order in perturbation (and also found at the mean-field level) is not modified. This is expected as the calculation involves only spin excitations with no momentum change. On the contrary, the renormalization of the two-body interaction and the emergence of a three-body interaction are modified. In the simple case where $g_6=0$:
\begin{align}
E^{(2)}&=-\frac{(g^\textrm{1D}_4)^2N^2 \sqrt{m}}{4L\sqrt{2\hbar\tilde{\Omega}}\hbar},\\
E^{(3)}&=\frac{(g^\textrm{1D}_4)^2g^\textrm{1D}_3 N^3}{ 32\sqrt{2}L^2}\frac{1}{(\hbar\tilde{\Omega})^{3/2}}\frac{\sqrt{m}}{\hbar},
\end{align}
where $L$ is the size of the 1D system. The sign of the dominant two-body term is now negative, in contrast to the 3D case. This may be expected since for $\Omega=0$, the beyond-mean-field correction is negative in 1D \cite{Petrov2016}. In addition, the scaling with $\Omega$ is modified as compared to the 3D case. These corrections can also be found in a beyond-mean-field approach in the presence of a coupling similarly to what was done in 3D \cite{Lavoine2021}. 

\section{Conclusion}
In this paper, we have performed a perturbative treatment of the interactions in a Bose dressed system with competing interactions. This has been done not only for the energy of a condensate but also for the two-body and three-body scattering problems. Our approach permits us to nicely explain the origin of the different kinds of three-body interactions in a unified framework at different orders in perturbation theory.  It justifies the use of an effective Hamiltonian that includes a contact three-body interaction term when conditions for a valid perturbative expansion are met. The control of three-body interactions through radio-frequency-dressing appears as an important addition to the atomic gas toolbox. Precise calculations of the values of $g_{2b}$ and $g_{3b}$ for specific experimental conditions ($a_{\up \up}$, $a_{\up\dow}$, ,$a_{\dow \dow}$, $\delta$, $\Omega$) are straightforward. Interestingly, losses and short-range physics corrections would appear only at higher-order order in the expansion and are thus expected to be negligible compared to the three-body interactions found in the present work.

For compensating two-body interactions (with scattering lengths of opposite signs between the different spins), the dressing permits one to reach situations where the two-body interactions are reduced, such that higher-order three-body interaction terms are sizable and may even dominate the equation of state. This is in contrast to single-component dilute Bose gases where three-body interactions are usually negligible and/or associated with losses when the scattering length is increased \cite{Mestrom19}. Previous experimental works on three-body interactions in dressed Bose gases have focused on the modification of the equation of states \cite{Lavoine2021, Hammond2022}. Future works may directly detect the specific collision halos that would originate from three-body collisions. This could be done by studying the collision between two condensates under conditions such that there are no two-body collisions ($g_{2b}=0$) but only three-body collisions ($g_{3b}=0$).  

\begin{acknowledgments}
 This  research  was  supported  by  CNRS,  Minist\`ere  de  l'Enseignement  Sup\'erieur  et  de  la  Recherche,  Quantum Paris-Saclay, R\'egion Ile-de-France  in  the  framework  of  Domaine d'Int\'er\^et Majeur Quantip (H3B), France 2030 in the framework of PEPR Quantique (ANR-23-PETQ-0001), the Simons Foundation (Award No. 563916,  localization of waves), France and Chicago collaborating in the sciences.
\end{acknowledgments}



\bibliography{references}

\end{document}